\documentclass{article}
\usepackage{graphicx}

\usepackage{amssymb,amsthm,amsmath}
\usepackage{xcolor,paralist,hyperref,titlesec,fancyhdr,etoolbox}
\newtheorem{theorem}{Theorem}[]

\newtheorem{lemma}[theorem]{Lemma}

\newtheorem{corollary}[theorem]{Corollary}

\newtheorem{problem}[theorem]{Problem}
\usepackage{lipsum}

\usepackage{booktabs} 
\usepackage{xcolor}

\usepackage{floatflt}
\usepackage{url}

\usepackage{algcompatible}
\usepackage{algorithm}
\usepackage{algpseudocode}

\begin{document}
\title{Specifying and Verifying the Convergence Stairs of the Collatz Program} 
\author{Ali~Ebnenasir \\ Department of Computer Science\\Michigan Technological University\\Houghton MI 49931 \\Email: aebnenas@mtu.edu\\}
\maketitle


\begin{abstract}
This paper presents an algorithmic method that, given a positive integer $j$, generates the $j$-th convergence stair containing all  natural numbers from where the Collatz conjecture holds by exactly $j$ applications of the Collatz function. To this end,  we present a novel formulation of the Collatz conjecture as a concurrent program, and provide the general case specification of the $j$-th convergence stair for any $j > 0$. The proposed specifications provide a layered and linearized orientation of Collatz numbers organized in an infinite set of infinite binary trees. To the best of our knowledge, this is the first time that such a general specification is provided, which can have significant applications in analyzing and testing the behaviors of complex non-linear systems. We have implemented this method as a software tool that generates the Collatz numbers of individual stairs. We also show that starting from any value in any convergence stair the conjecture holds. However, to prove the conjecture, one has to show that every natural number will appear in some stair; i.e., the union of all stairs is equal to the set of natural numbers, which remains an open problem.

\end{abstract}


%
%
%

\vspace*{-5mm}
\section{Introduction}
\label{sec:Intro}
\vspace*{-2mm}

The objective of this paper is to present a systematic approach for the {\em specification} of Collatz numbers based on their distance to the set of powers of two. 
Consider the function $f_c$ over a variable $x$ whose domain and range include the set of positive integers, denoted $\mathcal{N}$ : if $x$ is an even value, then update $x$ with $f_c(x) = x/2$; otherwise, assign $f_c(x) = 3x + 1$ to $x$. Thus, starting from a positive integer $n$, one can define a sequence of values $n, f_c(n), f_c^2(n), f_c^3(n), \cdots $ obtained by repetitive application of the Collatz function $f_c$, called the {\it orbit} of $n$. Let $f_{min}(n)$ be the minimum value in $\{n, f_c(n), f_c^2(n), f_c^3(n), \cdots \}$.  The Collatz conjecture simply asks the question of whether $f_{min}(n) = 1$ for any positive integer $n$. 

The significance of the  Collatz conjecture lies in its applications in several domains such as
image encryption \cite{ballesteros2018novel}, software watermarking \cite{ma2019xmark}, and in designing the stability of complex and non-linear systems \cite{grauer2021analogy} that have chaotic behaviors, yet ensuring  eventual stability. The notion of eventual stability is common knowledge in Computer Science in general, and in the self-stabilization community in particular, however, the behaviors of Collatz function defy any known type of convergence-assurance approach (e.g., ranking functions \cite{itcsStomp93},
convergence stairs \cite{stabCom91,tseGouda02}) as we know it in self-stabilizing systems. Moreover, while there are other distributed programs with unbounded variables (e.g., Dijkstra’s
token passing \cite{dij}), the domain of such unbounded variables often increases as the
network size grows; i.e., unbounded but finite. This is not the case in Collatz function, and the domain of x is infinite in a two-process concurrent program. Thus, the Collatz problem poses an interesting challenge for formal methods and program verification research too. 

While there is a rich body of work in mathematics on Collatz problem, the most recent result states that “Almost all Collatz orbits attain almost bounded values.” \cite{tao2022almost}, where the notion of “almost all" is defined in the context of logarithmic density. We take a different approach by first formulating the Collatz function as a concurrent program (Section \ref{sec:colPrg}). Then, we reformulate the Collatz problem as a problem of specifying and verifying the convergence of the {\it Collatz programs} through the specification of an unbounded number of convergence stairs (Section \ref{sec:cnvrg}). Each {\it convergence stair} is in fact a set of natural values from where the set $I_{cltz} = \{1, 2, 4\}$ can be reached in $j > 0$ steps of applying the Collatz function $f_c$. Formally, the $j$-th convergence stair, denoted $S_j$, is equal to the set $\{n \mid f^j_c(n) \in  I_c\}$. (Notice that, $S_0 = I_{cltz}$.) Our objective is to devise a scheme where, given $j > 0$ one can compute all the values in $S_j$ without expanding and exploring the binary tree generated by backward reachability from $\{1, 2, 4\}$. 
This way, the Collatz conjecture would be reduced to proving that (1) $\cup_{j=0}^\infty S_j = \mathcal{N}$, and (2) from each stair $j>0$ the Collatz program reaches a value in stair $j-1$. Not only does this approach provide a different method of tackling the Collatz conjecture, but also it enables an algorithmic way for analyzing the behavior of every individual stair. Moreover, this approach provides a layered and linearized orientation of Collatz numbers organized in an infinite set of infinite binary trees. Such linearization methods can provide insight  in addressing other similar conjectures (e.g.,  Kakutani conjectures \cite{briscese2024conjectures}). 
 To the best of our knowledge, this is the first time that such a general specification is developed, which can have significant applications in understanding and testing of the behaviors of complex non-linear systems. For example,  designers can study how neighboring stairs interact. We study this method for convergence to $I_{cltz}$ and $\mathcal{I}_u= \{2^k \mid k \geq 0 \}$, and show that specifying and verifying convergence stairs for  $\mathcal{I}_u$ are tractable problems and more useful. Specifically, we present an algorithm that takes as input a value $j$ and generates all the values belonging to the $j$-th stair of converging to $\mathcal{I}_u$. We then show that every value generated is in fact a correct Collatz number in the $j$-th stair, and prove that our algorithm does not miss any value. The proof of correctness is performed through attaching a Binary Verification Code (BVC) to each number during its specification.  During the generation phase, we use the BVC to verify the correctness of the generated value. An implementation of the proposed algorithm is available at \url{https://github.com/aebne/CollatzStairs}.

\noindent{\bf Organization}.\ Section \ref{sec:colPrg} provides a characterization of the problem as a two-process concurrent program. Section \ref{sec:cnvrg} then investigates the specification and verification of convergence stairs. Section \ref{sec:related} discusses related work. Finally, Section \ref{sec:concl} makes concluding remarks and discusses some open problems.

\vspace*{-3mm}
\section{Collatz Program, Its Invariant and the Verification Problem}
\label{sec:colPrg}
\vspace*{-2mm}

Let $P_{cltz}$ denote the Collatz program. We refer to $x$ as the state variable of $P_{cltz}$ and the value of $x$ identifies the current state of $P_{cltz}$. Throughout this paper, we interchangeably use the terms ‘state’, ‘value', and `number'. In fact, each natural number represents a state of Program $P_{cltz}$. The $P_{cltz}$ program includes the following actions:


\vspace*{-1mm}
\begin{tabbing}
\hspace{5mm}\=$P_1:\hspace{2mm} (x \mod 2) = 0 $ \hspace{5mm}\=$ \rightarrow x := x/2$\\
\> $P_2:\hspace{2mm} (x \mod 2) \neq 0  $\>$\rightarrow  x := 3x+1$\\
\end{tabbing}
\vspace*{-4mm}

An action is a guarded command, denoted {\it grd}$\rightarrow${\it stmt}, where the  {\it grd} is a Boolean condition in terms of $x$ and the {\it stmt} specifies how $x$ can be updated when the guard holds; i.e., the action is {\it enabled}. Each action belongs to a separate process, namely $P_1$ and $P_2$. 
Notice that, there are no distribution constraints as the two processes can atomically read and write the program state (i.e., value of $x$). A {\it computation} of $P_{cltz}$ is a sequence of integer values generated by actions $P_1$ and $P_2$ starting from any value in $\mathcal{N}$ .

\noindent{\bf Invariant, closure and convergence}.\  Note that starting in any state in the set $I_{cltz} = \{1, 2, 4\}$, the
computations of $P_{cltz}$ remain in $I_{cltz}$; i.e., {\it closure}. A weaker version of $I_{cltz}$
is $\mathcal{I}_u = \{2^k \mid k \in \mathcal{N} \wedge k \geq 0\}$ because starting from $2^k$
(for $k \geq 0$), $\mathcal{I}_u$ remains closed in computations of $P_{cltz}$, and successive execution of action $P_1$ from any state in $\mathcal{I}_u$ would result in a state  in $\mathcal{I}_u$. We refer to $\mathcal{I}_u$ as the {\it unbounded} invariant of $P_{cltz}$. We define {\it convergence} to the invariant as the eventual reachability of the invariant by computations of $P_{cltz}$ from any arbitrary value in the set of natural numbers $\mathcal{N}$. In other words, the requirements of convergence to $\mathcal{I}_u$ can be stated as the following three properties: (1) {\it deadlock-freedom} outside $\mathcal{I}_u$: there is no value in $\mathcal{N} - \mathcal{I}_u$ where both actions of $P_{cltz}$ are disabled; (2)  {\it livelock-freedom} outside $\mathcal{I}_u$: there is no cycle $v_0, v_1, \cdots, v_k, v_0$ for $k \geq 0$ of values in $\mathcal{N} - \mathcal{I}_u$ such that $v_{i \oplus 1}$ can be obtained from $v_i$ by actions of $P_{cltz}$, where $0 \leq i \leq k$ and $\oplus$ denotes addition modulo $k+1$, and (3)  {\it divergence-freedom}: there is no value from where the computations of $P_{cltz}$ diverge to infinity. 

\noindent{\bf Self-stabilization}.\  A program $P$ is {\it self-stabilizing} to an invariant $I$ {\em iff} (if and only if) the invariant $I$ is closed in $P$, and starting from any state in $P$'s state space, the computations of $P$ converge to $I$. For instance, the Collatz conjecture, can be framed as the following problem:

 
 
 \begin{problem}
 \label{prob1}
 Does program $P_{cltz}$ self-stabilize to $\mathcal{I}_u$ from any value in $\mathcal{N}$? 
 \end{problem}

Notice that, convergence to $\mathcal{I}_u$ implies convergence to $I_{cltz}$ by actions of $P_{cltz}$.

\section{Convergence Stairs of the Collatz Program}
\label{sec:cnvrg}

This section investigates the problem of specifying convergence stairs for proving
the self-stabilization of the Collatz program $P_{cltz}$. Notice that, the Collatz program is deadlock-free outside $\mathcal{I}_u$ because any value is either even or odd, which would respectively enable either action $P_1$ or $P_2$. The convergence of $P_{cltz}$ to $\mathcal{I}_u$ is a hard problem
in part due to the infinite state space of $P_{cltz}$, and its chaotic behavior. In fact,
to ensure convergence, one has to show livelock-freedom and divergence-freedom of  $P_{cltz}$ in $\mathcal{N}- \mathcal{I}_u$. 
In this section, we study convergence to $I_{cltz}$ (Subsection \ref{sec:stair1}) and $\mathcal{I}_u$
(Subsection \ref{sec:stair2}) through the lenses of convergence stairs \cite{stabCom91,tseGouda02}. The $j$-th stair
includes the set of states from where an action of $P_{cltz}$ can take its state to some
state in the $(j-1)$-th stair, where $j > 0$. The stair zero includes the invariant
states. We note that our notion of a stair differs from that of \cite{stabCom91,tseGouda02} in that stairs
are disjoint sets of states (i.e., state predicates) in our work.

\subsection{Convergence Stairs With Respect to $I_{cltz}$}
 \label{sec:stair1} 
 
 We consider the orientation of states with respect to the number of steps it takes for $P_{cltz}$ to reach a state in $I_{cltz}$; called the {\it rank} or {\it stair} of a state. To this end, we perform a backward reachability analysis using the one-to-at-most-two relation $R(x)$ below, which computes the inverse of $f_c(x)$. 
 
\begin{equation}
{R}(x)=
    \begin{cases}
        2x & \\
        (x-1)/3 & \text{if } (x-1)/3  \    \text{is an odd integer; otherwise, undefined.}
    \end{cases}
\end{equation}

\begin{figure} [h]
\centering
\includegraphics[  scale= 0.12]{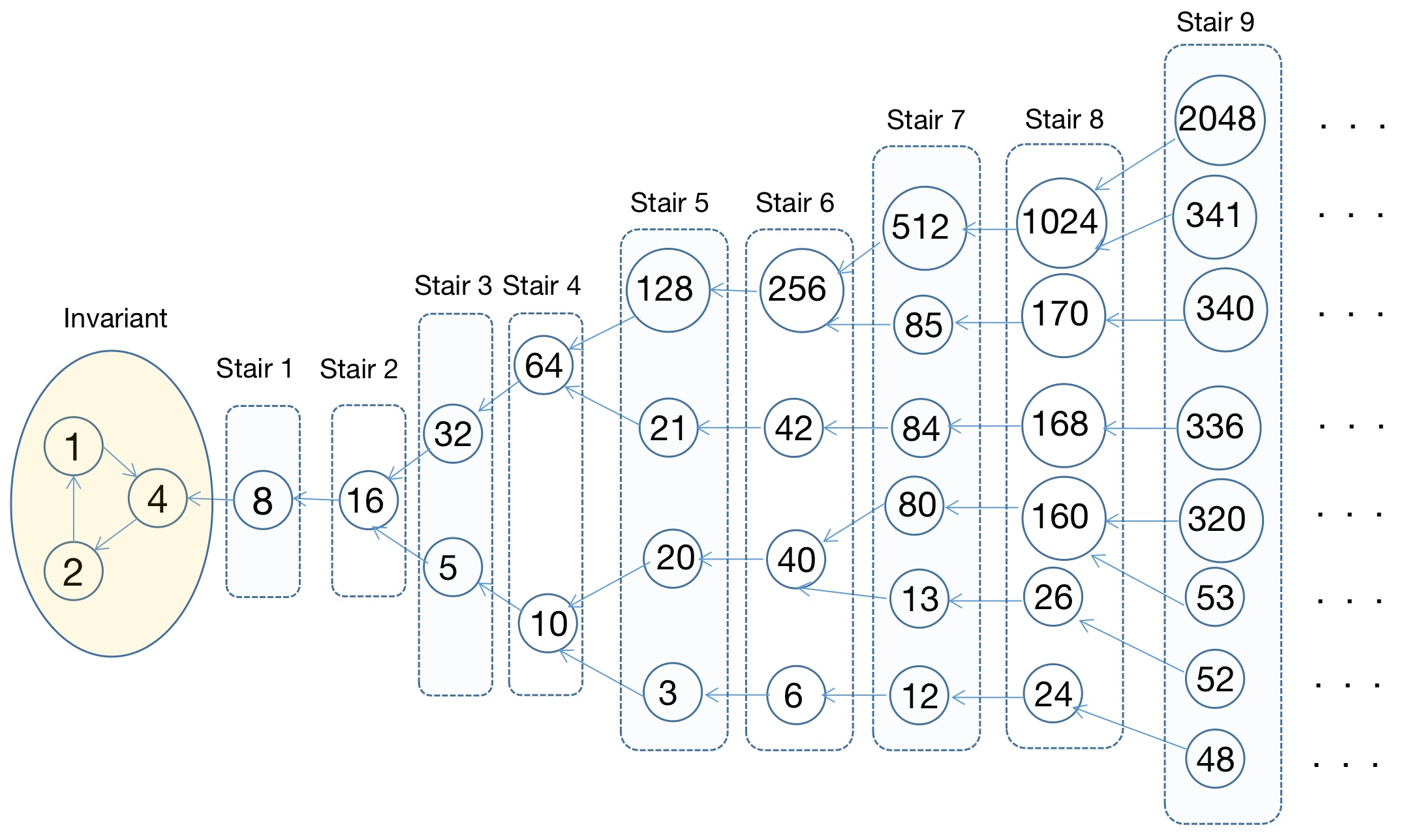}
\vspace*{-2mm}
\caption{\small Collatz computation tree and convergence stairs with respect to $I_{cltz}$.}
\label{fig:tree1}
\end{figure}

Expansion of $R(x)$ from 1 results in an infinite tree whose root includes the cycle $4 \rightarrow 2 \rightarrow 1 \rightarrow 4$. Figure \ref{fig:tree1} is a visualization of such a tree. This tree captures the computations of $P_{cltz}$ as an infinite binary tree, called the computation tree of $P_{cltz}$. We can consider each level of this tree as a  stair that converges to   the next lower level. The first stair contains $\{8\}$, the second stair is $\{16\}$,  the third stair is $\{5, 32\}$, the fourth stair has $\{10, 64\}$, and so on. 
This way of thinking about Problem \ref{prob1} may simplify the problem. The {\it orbit} of a state/value $n$ assigned to $x$ includes all values assigned starting from $n$ down to $1$. We call such an orbit the {\it recovery path} from $n$.
The $k$-th {\it stair}, where $k \geq 1$, includes all states from where there is exactly $k$ steps to $I_{cltz}$. 
 In this orientation of stairs, the largest value in the $k$-th stair is $2^{k+2}$. The $k$-th stair can also be represented through the application of the relation $R(x)$ $k$ times, denoted $R^{k}(x)$.
For example, we have $R^0(8)= \{8\},  R^1(8)= \{16\}, R^2(8)= \{5, 32\}, R^3(8)= \{10, 64\}$, and $R^4(8)= \{3, 20, 21, 128\}$. Notice that, convergence from $R^k(8)$ to $R^{k-1}(8)$ through the actions of $P_{cltz}$ is guaranteed, for any $k > 1$. 
 By a misuse of notation, we apply $R(x)$ to a set of states/values too. That is, $R(S) = \{ x' \mid  \exists x_0: x_0 \in S: x' = R(x_0) \}$. Figure \ref{fig:tree1} illustrates the convergence stairs for $I_{cltz}$  as an infinite tree whose root includes the states in $I_{cltz}$.  We prove the following lemmas on the structural correctness of the infinite tree in Figure \ref{fig:tree1}.

\begin{lemma}
\label{lma:stairs}
All values in stair $k+1$ are backward reachable by $ R(x)$ from all values in the $k$-th stair, where $k \geq 1$. That is, for any value $y$ in the $(k+1)$-th stair, there is some value $x$ in the $k$-th stair such that $y \in R(x) $. (Proof straightforward; hence omitted.)
\end{lemma}

\begin{lemma}
\label{lma:stairCmplt}
The set of values in the $k$-th stair is complete, for $k \geq 1$. That is, there is no positive integer $x_0$ from where $P_{cltz}$ can reach $I_{cltz}$ by exactly $k$ transitions/steps such that $x_0$ is missing in the set of values in the $k$-th stair.
\end{lemma}

\begin{proof}
We prove this lemma by induction on $k$ and the set of values  $R^k(I_{cltz}) $ generates.

\begin{itemize}
\item {\it Base case} (i.e., $k=1$): The only state in $\neg I_{cltz}$ from where $I_{cltz}$ is reachable by a single step of $P_{cltz}$ includes 8. This is also shown by computing $R(I_{cltz}) - I_{cltz} = \{1, 2, 4, 8\} - I_{cltz} = \{8\}$. Thus, there is no missing value $x_0$ in the first stair from where $P_{cltz}$ can reach $I_{cltz}$ by exactly $1$ transition.
 
\item  {\it Induction hypothesis}: There is no state $x_0$ missing from the $k$-th stair for $k \geq 1$ such that $I_{cltz}$ can be reached from $x_0$ by the actions of $P_{cltz}$ in exactly $k$ steps.

\item    {\it Inductive step}:  We show that the $(k+1)$-th stair is complete; i.e., no state $x_0$ is missing from the $(k+1)$-th stair for $k>1$. Since the $k$-th stair is complete, applying $R(x)$ to every single state in the $k$-th stair would result in all states from where the states in the $k$-th stair can be reached in a single step. This means that the $(k+1)$-th stair contains all states from where $I_{cltz}$ can be reached  by the actions of $P_{cltz}$ in $k+1$ steps. Therefore, the $(k+1)$-th stair is complete.
\end{itemize}
\end{proof}

Now, we state two important problems that are worth investigating (in our opinion).

\begin{problem}
\label{probComplt}
The set of backward reachable state from  $I_{cltz}$ is complete. Formally, $\{1,2,4\} \cup (\cup_{k=0}^{\infty} R^k(8)) = \mathcal{N}$. 
\end{problem}

Solving Problem \ref{probComplt} amounts to solving the Collatz conjecture. To prove the conjecture, one can prove that all positive integers can be generated by this backward reachability method; i.e., $I_{cltz} \cup (\cup_{k=0}^{\infty} R^k(8)) = \mathcal{N}$. In other words, any positive integer in $\neg I_{cltz}$ belongs to some stair.

\begin{problem}
\label{probSpec}
Design a function {\it stair}$: \mathcal{N} \rightarrow 2^\mathcal{N}$ that takes the index of a stair, and returns the set of states in that stair.
\end{problem}

This is an interesting specification problem from the computer science point of view, where {\it stair}$(k)$ is the specification of the $k$-th stair. Looking at Figure \ref{fig:tree1}, we observe that the states in {\it stair}$(k)$ have an upper bound of $2^{k+2}$ (in terms of their value), but it is unclear how one can specify {\it stair}$(k)$ so it accurately identifies all states of the $k$-th stair without performing a backward reachability analysis from $I_{cltz}$ using $R(x)$.

\begin{problem}
\label{idenStair}
Given a positive integer $n \in \neg I_{cltz}$, in which stair would $n$ be located? That is, design a function {\it stairIndex}$(n): \mathcal{N} \rightarrow \mathcal{N}$ that takes an integer $n$ and returns another integer  {\it stairIndex}$(n)$ which determines the number of steps required to reach $I_{cltz}$ by the actions of $P_{cltz}$ from $n$.
\end{problem} 

Solving Problem \ref{idenStair} will help in solving Problem \ref{probComplt} because one can use {\it stairIndex}$()$ to verify whether there is any integer $x_0$ that does not belong to any stair; i.e., {\it stairIndex}$(x_0)$ is undefined. We also observe that a solution to Problem \ref{probSpec} would have a similar impact since one can reason about $\cup_{k=1}^\infty${\it stair}$(k)$ and its equality to $\mathcal{N}-\{1,2,4\}$. However, addressing any one of the above questions seems to be as hard as solving the Collatz conjecture itself, but it is useful to think about the Collatz conjecture differently. For this reason, the next section provides an alternative method of forming the stairs towards solving Problem \ref{probSpec}.

\subsection{Convergence Stairs With Respect to $\mathcal{I}_u$}
 \label{sec:stair2} 
 
\begin{figure} [h]
\includegraphics[  scale= 0.15]{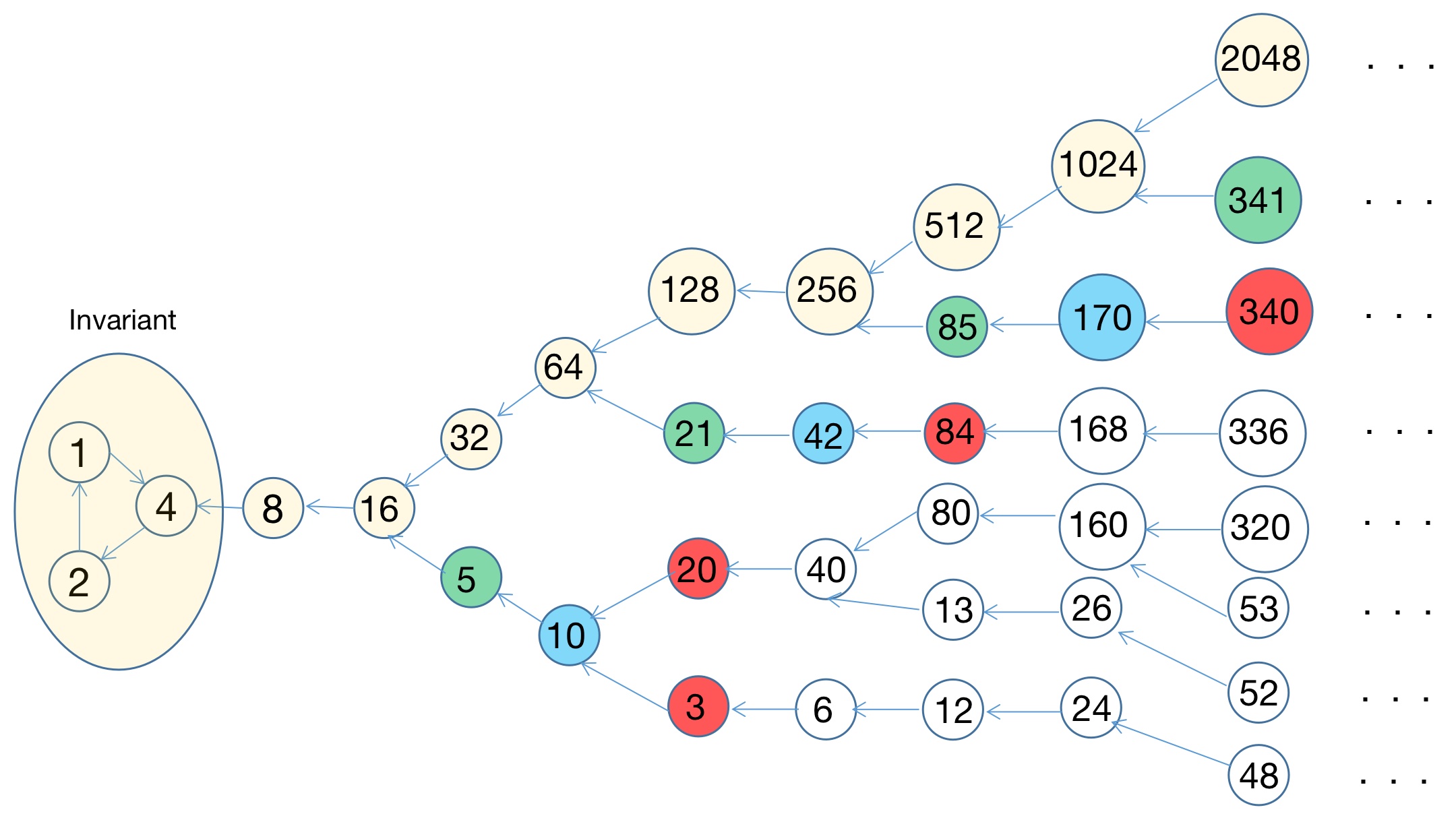}
\vspace*{-2mm}
\caption{\small Collatz computation tree and convergence stairs with respect to $\mathcal{I}_u = \{1, 2, 4\} \cup \{2^k \mid k \in \mathcal{N}  \wedge (k > 2) \}$. Green states represent Stair 1, Blue states capture Stair 2, and Red states illustrate Stair 3 with respect to $\mathcal{I}_u$.}
\label{fig:tree2}
\end{figure}

Figure \ref{fig:tree2} illustrates a different way of thinking about the infinite computation tree of the Collatz program, where the invariant is the unbounded set $\mathcal{I}_u$ instead of the finite invariant $I_{cltz}$. 
The first stair in Figure  \ref{fig:tree2} includes all the green states, which we can specify formally as $(2^{2k}-1)/3$, for $k>1$. The following lemma proves the correctness of this specification.

\begin{lemma}
\label{stOne}
$(2^{2k}-1)$ is divisible by 3 and $(2^{2k}-1)/3$ is an odd value, for $k > 0$.
\end{lemma}
 
 \begin{proof}
We prove this lemma by induction on $k$.

\begin{itemize}
\item {\it Base case}: For $k=1$, we have $(2^{2k}-1)= 3$, which is obviously divisible by 3 and $3/3$ is odd. For $k=2$, $(2^{2k}-1)= 15$ and $15/3$ is odd.

\item  {\it Induction hypothesis}: $(2^{2k}-1)$ is divisible by 3 and $(2^{2k}-1)/3$ is odd for some $k > 1$.

\item    {\it Inductive step}:  We show that for any $k>0$, if $(2^{2k}-1)$ is divisible by 3 and $(2^{2k}-1)/3$ is odd, then $(2^{2k+2}-1)$ is divisible by 3 and $(2^{2k+2}-1)/3$  is odd too. We rewrite $(2^{2k+2}-1)$ as $((2^{2k} \times 4)-1)$. Thus, we have \\
\hspace*{1cm} $(2^{2k} \times 4)-1 =(2^{2k} \times 4)+(-4 +3) =$\\
\hspace*{1cm} $ 4 (2^{2k}-1) + 3 =$ (By the hypothesis, the value inside parenthesis is divisible by three.) \\
\hspace*{1cm} $4 (3x) + 3 = 3 (4x+1)$ \\
 
The value $3 (4x+1)$ is a multiple of 3; hence divisible by three. Its division by three, i.e., $(4x+1)$, is an odd value too.  
\end{itemize}
\end{proof}

\begin{lemma}
\label{stTwo}
 $2^m-1$ is not divisible by 3 for odd values of $m > 1$.
\end{lemma}

\begin{proof}
We prove this lemma by induction on $m$. 

\begin{itemize}
\item {\it Base case}: For $m=3$, we have $(2^3-1) = 7$, which is not divisible by 3. Moreover, when $m=5$, we have $(2^5-1) = 31$, which is again not divisible by 3. 

\item  {\it Induction hypothesis}: $(2^m-1)$ is not divisible by 3 for  some odd value $m \geq  3$.

\item    {\it Inductive step}:  We show that, if $m'=m+2$ (i.e., $m'$ is the next odd value after $m$), then $(2^{m'}-1) $ is also not divisible by 3. We substitute $m'$ in $(2^{m'}-1) $, and we get $(2^{m+2}-1) = (2^m \times 2^2) -1 = (2^m \times 2^2) -4 + 3 = 2^2 \times (2^m -1) + 3$. If $2^2 \times (2^m -1) + 3$ were divisible by 3, then $2^2 \times (2^m -1) $ would be divisible by 3, which in turn would imply that $(2^m -1) $ would be divisible by 3. This is a contradiction with the induction hypothesis. 
\end{itemize}

 \end{proof}

Lemma \ref{stTwo} shows that the green children in Figure \ref{fig:tree2} only exists for even powers of 2; i.e., $2^4, 2^6, 2^8, \cdots$.

To simplify the specification problem, let $Y_k = (2^{2k}-1)$. Then, $Y_k/3$ is a
nice specification of the first stair (the green nodes in Figure \ref{fig:tree2}) that can give
us every single state in it, for any integer $k > 1$. For instance, for $k = 2$ we
have $Y_2/3 = (2^4-1)/3 = 5$, and $Y_3/3 = (2^6-1)/3 = 21$, and so on. This is
an achievement with respect to the first orientation of stairs (in Section \ref{sec:stair1})
because it solves Problem \ref{probSpec} for the first stair. Since the elements of the first
stair are odd values (by Lemma \ref{stOne}), applying the first rule of $R(x)$ would give
us $2Y_k/3$ as a subset of elements in the second stair. The second rule of $R(x)$
would subtract 1 from $Y_k/3$ and divide it by three.

\begin{lemma}
\label{oneChild}
 $(Y_k/3 - 1)/3$ is not a valid Collatz number.
\end{lemma}

\begin{proof}
Lemma \ref{stOne} shows that  $Y_k$ is divisible by three and  $Y_k/3$ is an odd value. Thus, $Y_k/3 -1$ is even. If $Y_k/3 -1$ is divisible by three, then dividing $Y_k/3 -1$ by three must give us an even value; otherwise, $Y_k/3 -1$ would have been odd. Thus,  $(Y_k/3 -1)/3 = (Y_k -3)/3^2$ is an even value. This means that the Collatz function $f_c$ would do a dive-by-two operation on $(Y_k -3)/3^2$ instead of applying the $3x+1$ rule; i.e., $f_c((Y_k -3)/3^2) \neq Y_k/3$. Therefore, starting from $Y_k/3$, the relation $R(x)$ would  give us only one child in the computation tree, and that is equal to $2Y_k/3$; i.e., $(Y_k/3 - 1)/3$ is not a valid Collatz number. In terms of the computation tree in Figure \ref{fig:tree2}, this lemma means that all green nodes have only a single blue child. 
 \end{proof}

\begin{figure} [h]
\includegraphics[  scale= 0.15]{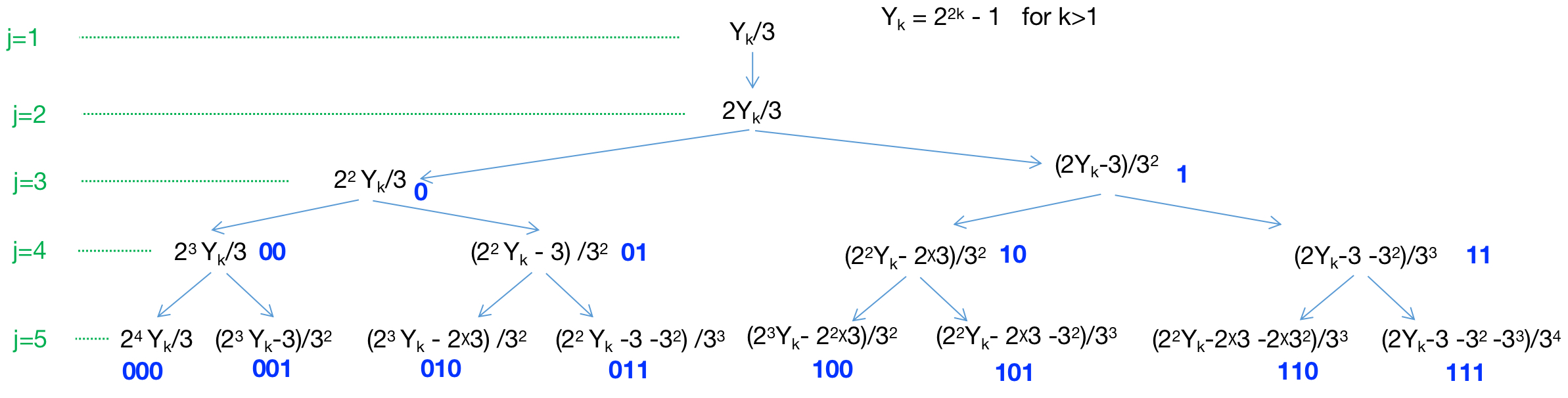}
\vspace*{-2mm}
\caption{\small Structure of a subtree rooted at $Y_k/3$, labelled with binary verification codes.}
\label{fig:annoTree}
\end{figure}

Based on Lemma \ref{oneChild}, the only members of the second stairs include  $2Y_k/3$ for $k > 1$ (see the Blue nodes in Figure \ref{fig:tree2}). Thus, we just solved Problem \ref{probSpec} for the second stair too. The specification of the third stair (Red states in Figure \ref{fig:tree2}) can be obtained by applying $R(x)$ on the blue states. The first subset of the third stair includes states $2^2Y_k/3$ due to applying the first rule of $R(x)$, and the second subset includes $(2Y_k -3)/3^2$. 
Continuing this way, in the fourth stair, we have the following values: $2^3Y_k/3,(2^2Y_k-3)/3^2,(2^2Y_k-2 \times 3)/3^2,(2Y_k-3-3^2)/3^3$. Figure \ref{fig:annoTree} illustrates the structure of the complete subtree rooted at $Y_k/3$ (for $k>1$) up to its fifth stair. Applying $R(x)$  further would give us an infinite binary  tree. Such a binary tree is an over-approximation because some of its nodes may not be valid Collatz numbers for the same reason stated in the proof of Lemma \ref{oneChild}. The main question is: {\it Given a specific $j>1$, how do we compute the Collatz numbers at the $j$-th level}?

Notice that, for each specific value of $k > 1$, we get a green value $Y_k/3$  in Figure  \ref{fig:tree2}, which is the root of a subtree by itself. Every stair $j > 1$ of a subtree rooted at $Y_k/3$ contains a single term $2^{j-1} Y_k/3$ corresponding to the left most branch of the tree rooted at $Y_k/3$. For example, for $j=4$ in Figure \ref{fig:annoTree}, we have a term $2^3 Y_k/3$. The remaining terms in the $j$-th stair are in the form of $(2^{f(j)}Y_k- (\Sigma^{q(j)-1}_{k=1} 2^{g(j)} \times 3^{h(j)}))/3^{q(j)}$, where $f(j) + q(j) = j$, $1 \leq h(j) \leq q(j)-1$ and $0 \leq g(j) \leq q(j)$ depending on the value of $q(j)$. For instance, if $j=1$, we have the value $Y_k/3$, where $f(j)=0$ and $q(j)=1$. 

\begin{lemma}
\label{lma:intrsct}
No two subtrees intersect.
\end{lemma}

\begin{proof}
By contradiction, consider two subtrees that intersect in a value $v$; i.e., $v$ has two parents, one in each tree. Thus, there are two distinct values $x_1 \neq x_2$ such that $R(x_1)=R(x_2) = v$. This means that $f_c(v) \in \{ x_1, x_2\}$, which is a contradiction with $f_c$ being a function.
\end{proof}

Now, the question is {\it how do we identify and specify  the functions $f(j), g(j), h(j), q(j)$ for values in each stair $j$ and some $k > 1$}? 

\begin{lemma}
For every term in the $j$-th stair, $f(j) + q(j) = j$ holds.
\end{lemma}

\begin{proof}
We show this by induction on $j$. 

\begin{itemize}
\item {\it Base case}: For $j=1$, we have the term $Y_k/3$, where $f(j)=0$ and $q(j)=1$. For $j=2$, $f(j)=1$ and $q(j)=1$ in $2Y_k/3$. That is, in both cases $f(j) + q(j) = j$ holds. 

\item {\it Induction hypothesis}: Assume that for $j \geq 2$, $f(j) + q(j) = j$ holds in  $(2^{f(j)}Y_k- (\Sigma^{q(j)-1}_{k=1} 2^{g(j)} \times 3^{h(j)}))/3^{q(j)}$.

\item  {\it Inductive step}: Progress to the next stair is made through two operations of $R(x)$: multiply by two, and subtract one and divide by three. If a child in the tree is obtained from $(2^{f(j)}Y_k- (\Sigma^{q(j)-1}_{k=1} 2^{g(j)} \times 3^{h(j)}))/3^{q(j)}$ by a multiply-by-two operation, then we have $f(j+1) = f(j)+1$ and $q(j+1) = q(j)$. Thus,  $f(j+1) + q(j+1) = f(j) + q(j) + 1$. By the hypothesis, we know $f(j) + q(j) = j$. Thus,  $f(j+1) + q(j+1) = j+1$. In the second case, a subtraction by one and division by three is performed. This operation does not increase $f(j)$; i.e., $f(j+1) = f(j)$. However, a division by three would increase $q(j)$ by one unit. That is, $q(j+1) = q(j) +1$. As such, $f(j+1) + q(j+1) = f(j) + q(j) +1$. Again, using the hypothesis, we have $f(j+1) + q(j+1) = j+1$.
\end{itemize}
\end{proof}

For each $j$, $f(j)$ start with $j-1$ and decrements down to 1. In turn, $q(j)$ starts at 1 and increments to $j-1$. That is, $1 \leq q(j) \leq j-1$; i.e., $q(j)$ belongs to the set $\{1, 2, \cdots, j-1\}$.  We now analyze the specification of the members of the $j$-th stair for different values of $q(j)$. This analysis will be the basis for designing an algorithm that takes $j$ and $k$, and generates the members of the $j$-th stair in the subtree rooted at  $Y_k/3$. Simultaneously, we present a scheme for verifying whether each individual term in the $j$-th stair is actually a valid Collatz number. The reason behind this is that the subtree rooted at $Y_k/3$ is an over-approximation, and not all terms have acceptable values based on the Collatz functions. For instance, in the absence of such verification, the subtree rooted at 5 in Figure \ref{fig:tree2} would include 4 as a child of 13 too, which is incorrect because $f_c(4)=2$. To enable such verification, we attach a binary string, called the {\it Binary Verification Code} (BVC), to each node of the tree, where a `0' indicates multiplication by two, and `1' represents subtraction of one and division by three. For example, the term $(2^2 Y_k - 2 \times 3 - 3^2)/3^3$ is obtained by the following three operations performed on $2Y_k/3$: (1) subtract one from $2Y_k/3$ and divide it by three, which results in $(2Y_k -3)/3^2$; (2) multiply by two, which gives $(2^2Y_k - 2 \times 3)/3^2$, and (3) finally, subtract one from $(2^2Y_k - 2 \times 3)/3^2$ and divide it by three, hence  $(2^2 Y_k - 2 \times 3 - 3^2)/3^3$. Another way to think about this is to attach 0 or 1 to the right side of the BVC of some node $v$ if you respectively go to the left or right child of  $v$. For instance, to reach $(2^2 Y_k - 2 \times 3 - 3^2)/3^3$ from $2Y_k/3$, we go right-left-right; hence the binary code 101 for $(2^2 Y_k - 2 \times 3 - 3^2)/3^3$. Now,  imagine we start with $(2^2 Y_k - 2 \times 3 - 3^2)/3^3$ and would like to verify whether it is an acceptable Collatz number. We scan its BVC from right to left. If we observe a `1', we multiply the current term by three and add one unit to derive its parent. Otherwise, we have a `0' and we divide the current term by two, and generate its parent. Then, consider  the derived parent as the current term and repeat scanning until all bits are processed. In each step, we can verify whether the child and the parent nodes meet the constraints of $f_c$. We note that, the notion of BVC is similar to the concept of parity vectors/sequences \cite{Everett1977} (also defined  in \cite{leventides2021koopman}).
In the following analysis, we discuss the BVCs of the generated terms in {\it italic}.

\begin{itemize}

\item  For $q(j)=1$, we have only one term  $2^{j-1}Y_k /3$ in the $j$-th stair. Since $q(j)=1$, we have $f(j)=j-1$ due to $f(j)+q(j)=j$. (That is why 2 is raised to $j-1$.)   {\it The corresponding BVC is the string $\langle 00 \cdots 0 \rangle$ of length $j-2$}. (See Lines 3-13 in Algorithm \ref{alg:stairAlg}) \\

\item For terms with $q(j)=j-1$ where $j>1$, we have one term $(2Y_k - (\Sigma_{i=1}^{j-2} 3^i))/3^{j-1}$ in the $j$-th stair. {\it  We have one single BVC $\langle 11 \cdots 1 \rangle$ of length $j-2$ bits.}   (See Lines 14-24 in Algorithm \ref{alg:stairAlg})\\

\item For $q(j)=2$ where $j>2$, we have $j-2$ terms in the $j$-th stair as follows: $(2^{j-2}Y_k - 2^{i} \times 3) /3^2$ for $0 \leq i < j-2$. {\it For each one of these terms, we have the following BVCs: $\langle 0 \cdots 1 \rangle, \langle 0 \cdots 10 \rangle, \langle 0 \cdots 100 \rangle, \cdots, \langle 10 \cdots 0 \rangle$ each of length $j-2$ bits.}  (See Lines 25-42 in Algorithm \ref{alg:stairAlg})\\

\item For terms with $q(j)=j-2$ where $j>2$, we have $j-2$ terms in the $j$-th stair, each of the form $(2^2Y_k - (\Sigma_{i=1}^{m} 2 \times 3^i + \Sigma_{i=m+1}^{j-3}  3^i ))/3^{j-2}$ where $0 \leq m \leq j-3$. Note that, in this case, $j-3 = q(j)-1$. {\it The corresponding  $j-2$ BVCs include $\langle 01 \cdots 1 \rangle, \langle 101 \cdots 1 \rangle, \langle 1101 \cdots 1 \rangle, \cdots, \langle 11 \cdots 01 \rangle, \langle 11 \cdots 10 \rangle$ of length $j-2$ bits each. In all these strings, there is a single 0 that moves from msb to lsb. }  (See Lines 43-68 in Algorithm \ref{alg:stairAlg})\\

\item For  $2 < q(j) < j-2$ where $j>5$,  the general form of the expressions includes $(2^{j - q(j)} Y_k - ( \Sigma_{k=1}^{q(j)-1}  2^i 3^k ))/3^{q(j)}$, where the sequence of exponents in powers of two is a non-increasing sequence out of the space of $(j-q(j)-1)^{q(j)-1}$ possible sequences. The maximum value of any exponent in powers of two is $(j-q(j)-1)$ and there are $q(j)-1$ terms in $\Sigma_{k=1}^{q(j)-1}  2^i 3^k $. We use Algorithm \ref{alg:recFor}  (invoked in Line 70 of Algorithm \ref{alg:stairAlg}) to compute such sequences recursively. The recursive nature of Algorithm \ref{alg:recFor} is due to the fact that the number of terms in $ \Sigma_{k=1}^{q(j)-1}  2^i 3^k $ depends on  $q(j)$ and their formation depends upon the exponents of powers of two. These two parameters both change from one stair to another. Thus, we need an algorithmic structure that dynamically changes; hence the recursion.  Algorithm \ref{alg:recFor} has six parameters. The first one captures the number of terms in $\Sigma_{k=1}^{q(j)-1}  2^i 3^k $; i.e., $q(j)-1$, the second parameter holds the largest possible exponent of two, the third one is a copy of the first one, and the remaining three parameters $j,qj, Y_k$ are passed for the calculation of the expression  $(2^{j - q(j)} Y_k - ( \Sigma_{k=1}^{q(j)-1}  2^i 3^k ))/3^{q(j)}$. The core of the algorithm includes Lines 2 to 5 where a nested for-loop is dynamically formed with the depth of $l$. During each recursion, we insert a value $0 \leq x  \leq l$ into a vector array, denoted {\it list}, which holds a permutation (with repetition) of $n=q(j)-1$ values in the domain $[0, ..., l]$. Algorithm \ref{alg:recFor} returns only those permutations that are non-increasing (stored in {\it list}[]).  In Line 18, Algorithm \ref{alg:recFor} uses the contents of {\it list}[] to compute $Num = (2^{j - q(j)} Y_k - (\Sigma_{r=1,t=list[r-1]}^{q(j)-1}  2^t \times 3^r ))/3^{q(j)}$. The remaining lines then use Algorithm \ref{alg:verif} to verify the acceptability of $Num$ as a Collatz number. 
\end{itemize}

Algorithm \ref{alg:verif} verifies whether a value $x$ is a legitimate Collatz number with the help of the binary string $s$ as the BVC of $x$. Initially, Algorithm \ref{alg:verif} performs some sanity checks (in Lines 1 to 6) to ensure that $x$ is an integer greater than 1 and not a power of 2. Then, it simply scans $s$ from its least significant bit (lsb). If the current bit (i.e., $s[i]$) is 1, then Algorithm \ref{alg:verif} calculates the parent of $x$ in the tree, denoted $y$, in Line 9; otherwise, it computes $y$ in Line 11. The condition $s[i]=1$ means that the last operation performed for the generation of $x$ was a subtraction by one and division by three. Thus, to derive the value from which $x$ was generated by $R$ (i.e., parent of $x$ in the tree), Algorithm \ref{alg:verif} performs the inverse in Line 9. Likewise for the case where  $s[i]=0$, Algorithm \ref{alg:verif} executes Line 11. Subsequently, Algorithm \ref{alg:verif} verifies whether $y$ is an acceptable natural value (in Lines 13-16). Then, in Lines 17-25, Algorithm \ref{alg:verif} checks the scenarios where the rules of $R(x)$ are applied incorrectly. When  $s[i]=1$, the value of $x$ cannot be even because the $3x+1$ rule has been applied to derive its parent. Moreover, if $s[i]=1$, then $x$ and $y$ cannot both be odd. When  $s[i]=0$, $x$ cannot be odd because the $x/2$ rule is applied (under $f_c$) to derive $y$. Then, in Line 30, $y$ becomes the current node/value in the tree for which verification must be done.

\begin{theorem}
Algorithm \ref{alg:verif} is correct. That is, it does not provide false positives nor does it output false negatives.
\end{theorem}

\begin{proof}
Algorithm \ref{alg:verif} returns True only when $x$ meets the following constraints: (1) $x$ is a natural value that is not a power of two, hence greater than 1 ($1=2^0$); (2) if the parent of $x$, denoted $y$, is generated by the rule $3x+1$ (i.e., $s[i]=1$), and $y$ is a natural value that is not a power of two, then it is not the case that $x$ is even. The reason behind this is that if $x$ where even, the rule $3x+1$ would not even apply under the Collatz function $f_c$.  Moreover, $x$ and $y$ cannot both be odd when the rule $3x+1$ is applied, and (3)  the rule $x/2$ is applied (i.e., $s[i]=0$) only when $x$ is even. Thus, when Algorithm \ref{alg:verif} returns true, $x$ has correctly been generated from the root of its subtree (i.e., $Y_k/3$) by successive application of $R$. Otherwise, Algorithm \ref{alg:verif} returns false if there is at least one step during the derivation of $x$ from $Y_k/3$ that does not meet the constraints of $f_c$. 
\end{proof}

\begin{theorem}
The asymptotic time complexity of Algorithm \ref{alg:verif} is  linear in the length of $s$.
\end{theorem}

\begin{proof}
The for-loop in Line 7 iterates $len(s)$ times. The asymptotic cost of the loop body is $O(1)$. Therefore, the asymptotic complexity of Algorithm \ref{alg:verif} is $O(len(s))$.
\end{proof}

\begin{algorithm}
\caption{The verification algorithm.}
\label{alg:verif}
\begin{algorithmic}[1]
\Require $x$ is a real value, $s$ is a non-null binary string.
\Ensure return a Boolean value

\If{$(x < 1) \; \vee $ ($x$ is not an integer)}
	\State return False;
\EndIf
\If{($x$ is a power of 2)} \Comment{{\it Internal values in a subtree cannot be powers of 2.}}
	\State return False;
\EndIf
\For{$i$ from $len(s)$ downto 1} \Comment{{\it Scan $s$ from right to left.}}
	\If{$s[i] = 1$}     \Comment{{\it Calculate the parent of $x$, i.e., $y$, when the $i$-th bit is 1.}}
		\State $y \gets 3x + 1$;
	\Else 		\Comment{{\it Calculate the parent of $x$, i.e., $y$, when the $i$-th bit is 0.}}
		\State $y \gets x/2$;
	\EndIf
	\If{$y$ is an integer greater than or equal to 1}
		\If{$y$ is a power of 2}
			\State return False;
		\Else  \Comment{{\it Cases where the rules of $R(x)$ are applied incorrectly.}}
			\If{$(s[i]=1) \wedge  (x \mod 2 = 0)$} \Comment{{\it  $x$ cannot be even when $s[i]=1$.}}
				\State return False;  
			\EndIf
			\If{$(s[i]=1) \wedge  (x \mod 2 \neq 0) \wedge  (y \mod 2 \neq 0)$} \Comment{{\it  $x$ and $y$ cannot both be odd when $s[i]=1$.}}
				\State return False;  
			\EndIf
			\If{$(s[i]=0) \wedge  (x \mod 2 \neq 0)$} \Comment{{\it  $x$ cannot be odd when $s[i]=0$.}}
				\State return False;  
			\EndIf					
		\EndIf
	\Else 
		\State return False;
	\EndIf
	\State $x \gets y$;	
\EndFor
\State return True;
\end{algorithmic}
\end{algorithm}

Algorithm \ref{alg:stairAlg} presents a method for simultaneous calculation of the Collatz numbers in the $j$-th stair of the subtree rooted at $Y_k/3$. Trivially, one can invoke Algorithm \ref{alg:stairAlg}  for different values of $k>1$ and some fixed $j$, or vice versa. The outline of Algorithm \ref{alg:stairAlg} follows the structure of the aforementioned analysis, where the BVC code of every term is generated with it. As a result, the output of Algorithm \ref{alg:stairAlg}  includes all Collatz numbers for some $k> 1, j >0$.

\begin{theorem}[Soundness]
Algorithm \ref{alg:stairAlg} is sound. That is, Algorithm \ref{alg:stairAlg} correctly explores the members of each stair $j$, for $j>0$.
\end{theorem} 

\begin{proof}
We prove this theorem by induction on $j$ for each case in the body of the for-loop in Line 2 of Algorithm \ref{alg:stairAlg}. This way we show the correctness of the expressions describing the members of the $j$-th stair. Note that, $q(j)$ represents the exponent of the denominator in a member of the $j$-th stair, where $1 \leq q(j) \leq j-1$.

\begin{algorithm}
\caption{An algorithm for specifying the members of the $j$-th stair for a given value $k>1$, along with constructing their binary verification code.}
\label{alg:stairAlg}
\begin{algorithmic}[1]
\Require $k$ and $j$ are positive integers where $k > 1$ and $j > 0$
\State $Y_k \gets 2^{2k}-1$
\For{$1 \leq q_j \leq j-1$}
\If{$q_j=1$}
    \State $x \gets 2^{j-1} Y_k/3$
    \State bvc $\gets$ null  \Comment{{\it Construct the binary verification code (bvc) as a string.}}
    \For{$1 \leq t \leq j-2$}
		 $bvc \gets  $ concat(bvc, `0') \Comment{{\it Attach 0 to bvc from right.}}
	\EndFor \Comment{{\it End of bvc construction.}}
	\If{verify(x,bvc) = True}
		\State print(x,bvc); 
	\Else
		\State print("Invalid Collatz number: Verification failed!");
	\EndIf
\EndIf
\If{$q_j=j-1$}
    \State  $x \gets 2Y_k - (\Sigma_{i=1}^{j-2}  3^i)/3^{j-1}$
        \State bvc $\gets$ null  \Comment{{\it Construct the bvc.}}
    \For{$1 \leq t \leq j-2$}
		 $bvc \gets $ concat(bvc, `1') \Comment{{\it Attach 1 to bvc from right.}}
	\EndFor   \Comment{{\it End of bvc construction.}}
	\If{verify(x,bvc) = True}
		\State print(x,bvc); 
	\Else
		\State print("Invalid Collatz number: Verification failed!");
	\EndIf
\EndIf
\If{$q_j=2$}
    	\For{$0 \leq i < j-2$}
		 \State $x \gets (2^{j-2} Y_k - 3 \times 2^i)/3^2$
		\State bvc $\gets$ null   \Comment{{\it Construct the bvc.}}
    		\For{$0 \leq t < j-2$}
			\If{$(t=i)$} 
				\State bvc $\gets $ concat(`1',bvc) \Comment{{\it Attach 1 to bvc from left.}}
			\Else
				 \State bvc $\gets $ concat(`0',bvc) \Comment{{\it Attach 0 to bvc from left.}}
			\EndIf
		\EndFor  \Comment{{\it End of bvc construction.}}
	\If{verify(x,bvc) = True}
		\State print(x,bvc); 
	\Else
		\State print("Invalid Collatz number: Verification failed!");
	\EndIf
	\EndFor
\EndIf
  \algstore{myalg1}
\end{algorithmic}
\end{algorithm}

\begin{algorithm}
\begin{algorithmic}[1]
\algrestore{myalg1}
\If{$q_j=j-2$}
    	\For{$0 \leq m < j-2$}
		\State $x \gets (2^2 Y_k - (\Sigma_{i=1}^m 2 \times 3^i + \Sigma_{i=m+1}^{j-3} 3^i))/3^{j-2}$
	\If{$(m = 0)$}
		\State bvc $\gets$ null   \Comment{{\it Construct the bvc.}}
		\For{$1 \leq t < j-2$} 
			\State bvc $\gets$  concat(bvc, `1')
		\EndFor
		\State print bvc;  \Comment{{\it End of bvc construction.}}
	\Else
		\State bvc $\gets$ null   \Comment{{\it Construct the bvc.}}
		\For{$m+1 \leq t < j-2$} 
			\State bvc $\gets$  concat(`1',bvc)
		\EndFor
		\State bvc $\gets$  concat(`0',bvc)

		\For{$1 \leq t < m+1$} 
			\State bvc $\gets$ concat(`1',bvc)
		\EndFor	\Comment{{\it End of bvc construction.}}	
		\If{verify(x,bvc) = True}
			\State print(x,bvc); 
		\Else
			\State print("Invalid Collatz number: Verification failed!");
		\EndIf
	\EndIf
	\EndFor
\EndIf
\If{$2 < q_j < j-2$}
	\State recursiveFor($q_j-1,j-q_j-1, q_j-1, j, q_j, Y_k$)  \Comment{{\it Recursively compute the remaining numbers.}}
\EndIf
\EndFor
\end{algorithmic}
\end{algorithm}

\begin{itemize}
\item  For $q(j)=1$, we have only one term  $2^{j-1}Y_k /3$. 

\begin{itemize}
\item {\it Base case}:  For $j=1$, we have the root of the tree, which is $Y_k/3$.  

\item {\it Induction hypothesis}:  For $j \geq 1$, the $j$-th stair contains a single term  $2^{j-1}Y_k /3$. 

\item {\it Inductive step}: The $(j+1)$-th stair also has a single term generated by multiplying  $2^{j-1}Y_k /3$ by two. Since  $2^{j-1}Y_k /3$ is the only term in the $j$-th stair with a denominator 3,  $2^jY_k /3$ is  also the only term of this form in the $(j+1)$-th stair.
\end{itemize}

\item For terms with $q(j)=j-1$ in the $j$-th stair where $j>1$, we have one term $(2Y_k - (\Sigma_{i=1}^{j-2} 3^i))/3^{j-1}$. 

\begin{itemize}
\item {\it Base case}:  When $j=2$, we have only the term $2Y_k/3$, which can be generated from $(2Y_k - (\Sigma_{i=1}^{j-2} 3^i))/3^{j-1}$ by substituting $j$ with 2.

\item {\it Induction hypothesis}:  For $j \geq 2$, the $j$-th stair includes a single term  $(2Y_k - (\Sigma_{i=1}^{j-2} 3^i))/3^{j-1}$, where $q(j)=j-1$

\item {\it Inductive step}: The term  $(2Y_k - (\Sigma_{i=1}^{j-2} 3^i))/3^{j-1}$ is in fact the right most child in the $j$-th level of the tree derived by successive application of the inverse of the $3x+1$ action from the term $2Y_k/3$. That is, subtract one unit and divide by three. As such, when we subtract one from  $(2Y_k - (\Sigma_{i=1}^{j-2} 3^i))/3^{j-1}$ and divide it by three, we get the child  $(2Y_k - (\Sigma_{i=1}^{j-1} 3^i))/3^j$ in the  $(j+1)$-th stair of the tree. 
\end{itemize}

\item For $q(j)=2$ and $j>2$, we have $j-2$ terms as follows: $(2^{j-2}Y_k - 2^{i} \times 3) /3^2$ for $0 \leq i < j-2$. 

\begin{itemize}
\item {\it Base case}:  For $j=3$, we have the term $(2Y_k - 3)/3^2$ in the third stair (see Figure \ref{fig:annoTree}).

\item {\it Induction hypothesis}:   For $j \geq 3$, we assume that there are $j-2$ terms of the form $(2^{j-2}Y_k - 2^{i} \times 3) /3^2$  in the $j$-th stair. 

\item {\it Inductive step}: Moving from the $j$-th stair to the $(j+1)$-th stair, each one of the $j-2$ terms in the $j$-th stair is multiplied by 2 and generates another term of the form  $(2^{j_u-2}Y_k - 2^{i_u} \times 3) /3^2$ in the $(j+1)$-th stair, where $j_u = j+1$ and $i_u = i+1$. There is one more term of this form that can be derived from $2^{j-1}Y_k/3$ by performing a subtraction of one and division by three. Therefore, in the $(j+1)$-th stair, we have $(j+1)-2$ terms of the form $(2^{(j+1)-2}Y_k - 2^{i} \times 3) /3^2$, where $0 \leq i < (j+1)-2$.
\end{itemize}

\item For terms with $q(j)=j-2$ and $j>2$,  we have $j-2$ terms, each of the form $(2^2Y_k - (\Sigma_{i=1}^{m} 2 \times 3^i + \Sigma_{i=m+1}^{j-3}  3^i ))/3^{j-2}$ where $0 \leq m \leq j-3$.

\begin{itemize}
\item {\it Base case}:  When $j=3$, we have $m = 0$. The term $(2^2Y_k - (\Sigma_{i=1}^{m} 2 \times 3^i + \Sigma_{i=m+1}^{j-3}  3^i ))/3^{j-2}$ reduces to a single term $2^2Y_k/3$ (see Figure \ref{fig:annoTree}).

\item {\it Induction hypothesis}:  For $j \geq 3$, we have $j-2$ terms in the $j$-th stair, each generated from $(2^2Y_k - (\Sigma_{i=1}^{m} 2 \times 3^i + \Sigma_{i=m+1}^{j-3}  3^i ))/3^{j-2}$ for some $0 \leq m \leq j-3$.

\item {\it Inductive step}:  Such terms are derived from the members of the $j$-th stair in two ways: (1) apply the subtract-one-then-divide-by-three rule to $j-2$ members of the $j$-th stair starting with $2^2Y_k $ (given to us by the hypothesis), and (2) apply the multiply-by-two rule to the only member of the $j$-th stair of the form $(2Y_k - (\Sigma_{i=1}^{j-2} 3^i))/3^{j-1}$ (proved in the case where $q(j) = j-1$). This would give us $(j+1)-2$ terms of the from $(2^2Y_k - (\Sigma_{i=1}^{m} 2 \times 3^i + \Sigma_{i=m+1}^{j-2}  3^i ))/3^{j-1}$ for some $0 \leq m \leq j-2$.
\end{itemize}


\item For  $2 < q(j) < j-2$ and $j>5$,  we have a single term $(2^{j - q(j)} Y_k - ( \Sigma_{i=1}^{q(j)-1}  3^i ))/3^{q(j)}$. The proof of the single term $(2^{j - q(j)} Y_k - ( \Sigma_{i=1}^{q(j)-1}  3^i ))/3^{q(j)}$ is straightforward and similar to previous cases where a member is derived from the left most branch of the tree rooted at $Y_k/3$; hence omitted.  {\it The corresponding BVC is $\langle 00 \cdots 01 \cdots 1 \rangle$, where the number of 1s is equal to $q(j)-1$ and the number of 0s is $j-q(j)-1$}.

We compute the rest of the members of the set using the recursive function `recursiveFor' (Algorithm \ref{alg:recFor}) in Lines 69-72 of Algorithm \ref{alg:stairAlg}.

\begin{itemize}
\item {\it Base case}:  The base case for $j=6$, where $2 < q(j) < 4$; i.e., $q(j)=3$. In this case, we have a single term $(2^3 Y_k - 3 - 3^2)/3^3$, and the rest of the terms are obtained from Algorithm \ref{alg:recFor}. The input parameters $n$ and $l$ of Algorithm \ref{alg:recFor}  are respectively equal to $q(j)-1 = 2$ and $j-q(j)-1=2$. Algorithm \ref{alg:recFor} generates the following terms  
$(2^3Y_k -  3 - 3^2)/3^3, (2^3Y_k - 2 \times 3 - 3^2)/3^3, (2^3Y_k - 2^2 \times 3 - 3^2)/3^3, (2^3Y_k - 2 \times 3 - 2 \times 3^2)/3^3, (2^3Y_k - 2^2 \times 3 - 2 \times 3^2)/3^3, (2^3Y_k - 2^2 \times 3 - 2^2 \times 3^2)/3^3$. These correctly identify the members of the sixth stair where $q(j)=3$.


\begin{algorithm}
\caption{Algorithm recursiveFor implements a nested for-loop with variable depth $n$ for computing Collatz numbers of the $j$-th stair where $2 < qj < j-2$.}
\label{alg:recFor}
\begin{algorithmic}[1]
\Require $n$ represents the length of the summation $\Sigma_{r,s} 2^r3^s$ in $Num= (2^{j-q_j}Y_k - (\Sigma_{r,s} 2^r3^s)) / 3^{q_j}$, which is the same as the depth of the nested for-loop.

\Require $l$ denotes the maximum exponent for the powers of 2 in  $\Sigma_{r,s} 2^r3^s$ used in $Num$.
\Require $n_c$ is a copy of $n$. The values  $j, qj, Y_k$ are passed to this function for the computation of the members of the $j$-th stair.

\If{$(n>1)$} 
	\For{$x := l$ down to 0}
		\State list.push($x$)  \Comment{{\it Add $x$ to the end of the list.}}
		\State recursiveFor($n-1,l, n_c, j, qj, Y_k$)
		\State list.pop($n_c-n$) \Comment{{\it Return the element at position  $n_c-n$.}}
	\EndFor
\Else
	\For{$x := l$ down to 0}
		\State list.push($x$)
		\State accept = True
		\For{$i := 0$ to $n_c-1$}    \Comment{{\it Powers of 2 in $Num$ must be non-increasing in terms of their exponents.}}
			\If{(list[i] $<$ list[i+1])} 
			 	\State accept = False
				\State break
			\EndIf
		\EndFor
		\If{(accept == True)} 
			\State $Num = (2^{j - q(j)} Y_k - (\Sigma_{r=1,t=list[r-1]}^{q(j)-1}  2^t \times 3^r ))/3^{q(j)}$ \Comment{{\it Use the contents of  list[] as powers of two  in $Num$}}.
			\State Compute the corresponding bvc.
			\If{(verify($Num$,bvc) = True)}
				\State print(x,bvc); 
			\Else
				\State print("Invalid Collatz number: Verification failed!");
			\EndIf
		\EndIf
	\EndFor

\EndIf
\end{algorithmic}
\end{algorithm}

\item {\it Induction hypothesis}:   Algorithm \ref{alg:recFor}  correctly generates the members of the $j$-th stair for $2< q(j) < j-2$ where $j \geq 6$. 

\item {\it Inductive step}: Given the hypothesis, we show that Algorithm \ref{alg:recFor} correctly generates the members of the $(j+1)$-th stair for an arbitrary but fixed $q(j)$ in $2< q(j) < (j+1)-2$ where $j > 6$. Let $Num_j = (2^{j - q(j)} Y_k - (\Sigma_{r=1,t=list[r-1]}^{q(j)-1}  2^t \times 3^r ))/3^{q(j)}$ be an arbitrary expression/member generated in the $j$-th stair, where {\it list} denotes the array vector used in Algorithm \ref{alg:recFor}.  Note that, any member whose denominator is equal to $3^{q(j)}$ is generated from some member of the $j$-th stair in two ways: (1) apply the subtract-one-then-divide-by-three rule to any member of the $j$-th stair whose denominator is  $3^{q(j)-1}$, and (2) apply the multiply-by-two rule to any  member of the $j$-th stair whose denominator is  $3^{q(j)}$. 

If we apply the subtract-one-then-divide-by-three rule to $Num_j$, then we get $(2^{j - q(j)} Y_k - (\Sigma_{r=1,t=list[r-1]}^{q(j)-1}  2^t \times 3^r) - 3^{q(j)}/3^{q(j)+1}$ in the  $(j+1)$-th stair. This means that Algorithm \ref{alg:recFor} should generate a sequence of exponents of powers of two with the contents $list[0], list[1], \cdots list[q(j)-2], 0$ because the last power of two (i.e., $2^0=1$) is attached due to the new term $3^{q(j)}$. Such a sequence of values is  generated by Algorithm \ref{alg:recFor} because the first two parameters $n$ and $l$ passed to it for the members of the $j+1$-th stair whose denominator is $3^{q(j)+1}$ include $(q(j)+1)-1$ and $(j+1) - (q(j)+1) -1$. 

The application of the multiply-by-two rule to $Num_j$ would generate some members of the $(j+1)$-th stair with the form $(2^{j+1 - q(j)} Y_k - (\Sigma_{r=1,t=list[r-1]}^{q(j)-1}  2^{(t+1)} \times 3^r))/3^{q(j)}$. Such a sequence of values is  generated by Algorithm \ref{alg:recFor} when it is invoked with the parameters $n=q(j)-1$ and $l=j+1 - q(j) -1$. That is, the domain of  $l$ is increased by one. Therefore, Algorithm \ref{alg:recFor} correctly generates the members of the $(j+1)$-th stair for an arbitrary but fixed $q(j)$ in $2< q(j) < (j+1)-2$ where $j > 6$.
\end{itemize}
\end{itemize}
\end{proof}

\begin{theorem}[Completeness]
Algorithm \ref{alg:stairAlg} is complete. That is, Algorithm \ref{alg:stairAlg} explores all Collatz numbers (i.e., does not miss any number).
\end{theorem} 

\begin{proof}
We use induction to show that Algorithm \ref{alg:stairAlg} generates all Collatz numbers in every stair $j$, for $j>0$ and any $k>1$. First, we note that, based on Lemma \ref{stTwo}, every number $2^m$, where $m$ is even, has a child of the form $(2^m - 1)/3$. These values (i.e., $Y_k/3$ for $k>1$) are the roots of the subtrees for $k>1$. Moreover, Lemma \ref{lma:intrsct} shows that the subtrees do not share any nodes/values; i.e., each subtree is an independent infinite binary tree rooted at $Y_k/3$. Thus, we prove this theorem for an arbitrary value of $k>1$ and its subtree rooted at $Y_k/3$. 

\begin{itemize}
\item {\it Base case}: Lemma \ref{stTwo} shows that Algorithm \ref{alg:stairAlg} generate all members of the first stair; i.e., $j=1$ (Green nodes in Figure \ref{fig:tree2}). Such numbers include $Y_k/3$, for any $k>1$.

\item  {\it Induction hypothesis}: For $j \geq1$, Algorithm \ref{alg:stairAlg} explores and generates all Collatz numbers in the tree rooted at $Y_k/3$.

\item  {\it Inductive step}: We show that all numbers in the $(j+1)$-th stair are explored.  We apply $R(x)$ to every legitimate Collatz  number $y$ in the $j$-th stair to conduct a backward reachability process, where we generate the children $x_1$ and $x_2$ of $y$ in the $(j+1)$-th stair. Since the tree is binary, there are no other children that can be explored; i.e., there is no unexplored child of $y$. Algorithm \ref{alg:stairAlg}  uses Algorithm \ref{alg:verif}  and the BVC codes of  $x_1$ and $x_2$ to verify whether $x_1$ and $x_2$ are legitimate Collatz numbers. Since $y$ is a Collatz number,  $x_1$ and $x_2$ are legitimate Collatz numbers, if verified by Algorithm \ref{alg:verif}. Therefore, there is no $k>1$ whose corresponding subtree rooted at $Y_k/3$ misses any Collatz number reachable from $Y_k/3$ by successive application of $R(x)$. 
\end{itemize}

\end{proof}

\begin{theorem}[Complexity]
\label{thm:compl}
The asymptotic time complexity of Algorithm \ref{alg:stairAlg} for the $j$-th stair is $O(j^{j+1})$.
\end{theorem}

\begin{proof} 
The main for-loop in Line 2 of Algorithm \ref{alg:stairAlg} iterates at most $O(j)$ times. The asymptotic complexity of the body of the main for-loop is dominated by the complexity of the `recursiveFor' function (i.e., Algorithm \ref{alg:recFor}), invoked on Line 70. The worst case time complexity of `recursiveFor' is $l^n$ because it implements a nested for-loop structure of depth $n$ where each for-loop iterates $O(l)$ times. In one end of the interval $2 < q_j < j-2$  where $q_j=j-3$, the largest value for $n$ (i.e., $q_j-1$) is $j-4$.   In this case, we have $l = j - q_j-1 = j-(j-4) -1 = 3$. Thus, in this case, the asymptotic complexity of Algorithm \ref{alg:recFor} is $O(3^{j-4}) = O(3^j)$. In the other end of the interval $2 < q_j < j-2$  where $q_j=3$, we have $n=2$ and $l=j-q_j-1 = j-4$. Thus, in this case, the overall complexity is $O(j^2)$. In the middle of the range $2 < q_j < j-2$ where $q_j = j/2$, we have $n = j/2-1$ and $l = j-q_j-1 = j-j/2-1 = j/2-1$. In this case, the asymptotic time complexity is $O(j^j)$, which dominates the other two cases. Therefore, the worst case time complexity of Algorithm \ref{alg:recFor} is $O(j (j^j)) = O(j^{j+1})$. 
\end{proof}

\begin{theorem}
Every state in the $j$-th stair, where $j > 1$, of a subtree rooted at $Y_k/3$, for some $k>1$, will reach a state in the $(j-1)$-th stair of that tree through the execution of the Collatz program; i.e., applying the Collatz function $f_c$. 
\end{theorem}

\begin{proof}
We prove this lemma by induction on $j$ for an arbitrary $k>1$ and its corresponding subtree rooted at $Y_k/3$. 

\begin{itemize}
\item {\it Base case}: For $j=2$, we have the number $2Y_k/3$, which is an even value, and dividing it by 2 would give $Y_k/3$.

\item  {\it Induction hypothesis}: Each state in the $j$-th stair, for $j \geq 2$, reaches a state in the $(j-1)$-th stair through the execution of the Collatz program; i.e.,  by applying $f_c$.

\item  {\it Inductive step}: We show that from each state in the $(j+1)$-th stair, executing the Collatz program will result in a state in the $j$-th stair. Let $y$ be an arbitrary Collatz number in the $j$-th stair. We know that using the relation $R(y)$ for the expansion of the subtree rooted at $Y_k/3$ is a backward reachability process that explores the possibility of generating two Collatz numbers as children of $y$.  Let $x_1$ and $x_2$ denote the children of $y$ in the $(j+1)$-th stair of the tree rooted at $Y_k/3$. Without loss of generality, assume that both $x_1$ and $x_2$ are verified by Algorithm \ref{alg:verif}. Thus, starting at $x_1$ or $x_2$, the Collatz program reaches the state $y$ in the $j$-th stair. 
\end{itemize}

\end{proof}

\begin{corollary}
\label{cor:reach}
Every state in the $j$-th stair with respect to $\mathcal{I}_u$ will reach a state in the $(j-1)$-th stair through the execution of the Collatz program.
\end{corollary}

\begin{theorem}
\label{thm:reach}
Starting from any state/value in any subtree rooted at $Y_k/3$ for some arbitrary $k>1$, the Collatz program will eventually reach a state in $\mathcal{I}_u$, and will subsequently reach $I_{cltz}$. (Proof follows from Corollary \ref{cor:reach}.)
\end{theorem} 

\noindent{\bf Discussion}.\ Theorem \ref{thm:reach} shows that starting from any state in any subtree rooted at $Y_k/3$ for some arbitrary $k>1$ the Collats program does not diverge to infinity, nor does it get trapped in a non-progress cycle. This implies  the reachability of the set of powers of two, and then from there the reachability of the set $\{1,2,4\}$, which meets the requirement of the Collatz conjecture. However, to prove the conjecture, we have to show that the union of $\mathcal{I}_u$ and all the values in all stairs of all subtrees is equal to the set of natural numbers, which remains an open problem. That is, show that $\mathcal{N} = \mathcal{I}_u \cup (\cup_{j=1}^{\infty} (\cup_{k=2}^{\infty} S_{k,j}))$, where $S_{k,j}$ denotes the set of states in the $j$-th stair of the subtree rooted at $Y_k/3$ for $k>1$. One way to tackle this problem is to look for any natural value outside $\mathcal{I}_u$ that fails to be placed in any subtree. To this end, we should solve Problem \ref{idenStair}.

Since the time complexity of Algorithm \ref{alg:stairAlg} is exponential (based on Theorem \ref{thm:compl}), we discuss some potential optimizations one can make to enhance the efficiency of the algorithm. First, we observe that the computation of the values of the $j$-th stair has great potential for parallelization. Specifically, one can instantiate several instances of Algorithm \ref{alg:stairAlg} for distinct values of $j$ and $k$ in an embarrassingly parallel way. That is, the $j$-th stair of each subtree rooted at $Y_k/3$ can be computed totally independently on a separate machine. Second,  each case in the body of the main loop for different values of $q_j$ can also be computed in parallel as they do not depend on each other. For example, the case where $q_j = j-1$ can be executed in parallel with the case where $q_j = j-2$. Third, Algorithm \ref{alg:recFor} can be unrolled and implemented iteratively for fixed values of $j$ and $k$. Our recursive design of Algorithm \ref{alg:recFor} is mainly for presenting the dynamic nature of nested for-loop inside Algorithm \ref{alg:recFor}. This is not the only way that the exponents of two in Algorithm \ref{alg:recFor} can be computed.

We would also like to emphasize the application of Algorithm \ref{alg:stairAlg} in the reachability analysis of infinite-state systems where program variables have a domain equal to natural numbers. Moreover, our approach can be used for tackling other similar conjectures as well as for analyzing the convergence of complex non-linear systems to stability.  Another important application is in blockchain technology \cite{bocart2018inflation} where Collatz orbits (and respectively stairs) provide a pseudo-randomness used for generating proof-of-work.

\vspace*{-4mm}
\section{Related Work}
\label{sec:related}
\vspace*{-2mm}
 
There is a rich body of work on the Collatz conjecture, which can broadly be classified into theoretical, computational and representation in other domains (e.g., term rewriting or graph theory).  On the theoretical front, some mathematicians prove \cite{krasikov2003bounds,tao2022almost} weaker statements than the conjecture itself. For instance, Tao \cite{tao2022almost}  shows that “Almost all Collatz orbits attain almost bounded values.” Leventides and Poulios  \cite{leventides2021koopman} formulate the Collatz conjecture through bounded linear operators, and study the properties of these operators and their relation with the Collatz orbits. 

Computational methods investigate the limits of natural values that actually convergence to $I_{cltz}$ through running the Collatz program from initial values  to the extent their computational resources permit. For example, Lagarias \cite{lagarias10ultimate}  computationally verifies the conjecture for values up to $5.78 \times 10^{18}$. Barina \cite{barina2021convergence}  then improves this result by verifying the convergence of values up to $1.5 \times 2^{70}$ using both a single-threaded  and a parallel implementation.  Barghout {\it et al.}   \cite{barghout2023statistical} analyze the Collatz conjecture probabilistically and   reason that chances of not converging is low. Another class of computational methods focuses on searching for a livelock outside  $I_{cltz}$. For instance, Eliahou \cite{eliahou19933x} computationally verifies that there are no livelocks with a length up to $1.7 \times 10^7$. Due to the limited computational resources, these methods can verify only a finite scope of integers. 


Many existing methods reduce the Collatz conjecture to problems in other domains. For example, St{\'e}rin \cite{sterin2020binary} improves algorithmic methods for the representation of ancestors of any value $x$ in the Collatz graph as a regular expression reg$_k(x)$, which captures the set of binary representations of any ancestor $y$ of $x$ from where $x$ can be reached in $k$ application of the $3x+1$ rule. Fabio and Francesco study conjectures similar to Collatz's \cite{briscese2024conjectures}. Hernandez \cite{hernandez2023collatz} uses modular arithmetic to show that each orbit can be captured by a word in a regular language accepted by a DFA. However, their proof of convergence is not rigorous. Yolcu {\it et al.} \cite{yolcu2021automated} develop a term/string rewriting system that terminates if and only if the conjecture is true. This method provides an alternative way of reasoning about the conjecture. Or{\'u}s-Lacort and  Jouis \cite{orus2022analyzing} present a manual proof by induction, whereas there are methods \cite{furuta2022proof} that use theorem provers to mechanically verify the conjecture.  Rahn  {\it et al.} \cite{rahn2021algorithm} consider odd numbers between $2^n$ to $2^{n+1}$ in a complete binary tree over natural values (starting from 1) and define rules for coloring them depending on their status regarding convergence. An odd number that is not yet proven to converge is colored black. The proof of convergence is performed by an algorithm, called the Golden Automaton. An odd number that is proven to converge, but it can have unproven offsprings in the tree is colored gold. The blue odd numbers are those that are proven to converge and their offsprings have been colored gold. This way, one can reason about the convergence of odd numbers from one level of the binary tree to another level; i.e., linearize  the converging odd numbers. By contrast, the proposed notion of convergence stairs and Algorithm \ref{alg:stairAlg} provide an algorithmic method for exact generation of Collatz values in each stair as a precise method of linearization without actually generating the numbers in lower-level stairs.

\noindent{\bf Program Verification Methods}. \
We now discuss the applicability of existing methods for the verification of parameterized/unbounded programs, where the code of the program is parameterized in terms of the number of processes and/or domain size of variables. If the domain of $x$ were finite, one could utilize existing verification and synthesis methods \cite{kulkarni2000automating} to solve this problem in polynomial time in the size of its state space. Nonetheless, the state space of  $P_{cltz}$ is $\mathcal{N}$, and existing finite-state methods are not applicable. Even methods \cite{ebnenasir2019topology} that verify and synthesize self-stabilizing protocols with an unbounded number of processes (i.e., {\it parameterized programs}) fail because they mostly assume constant-space processes. 
The cutoff methods \cite{kaiser2010dynamic,abdulla2016parameterized} are of little help because the Collatz program has a fixed number of processes, instead the variable domain is infinite. 

Predicate abstraction \cite{graf1997construction,ball2001automatic} preserves the control flow structure of a program but creates an abstract version of the program with only Boolean variables representing data and control structures. Such abstractions are mostly useful for safety properties and local liveness properties, whereas in the case of Collatz program any finite abstraction of $\mathcal{N}$ must guarantee convergence from every single concrete state; i.e., predicate abstractions have little chance for simplifying reasoning about convergence. For example, Abdulla {\it et al.} \cite{abdulla2007parameterized} verify safety of infinite-state concurrent programs through creating an over-approximation of the transition function, and then conduct a backward reachability analysis, which may not terminate in general. Bultan {\it et al.} \cite{bultan1999model} encode state transition relations of unbounded concurrent programs as Presburger formulas, and use a Presburger solver to reason about their safety and liveness properties. However, their approach can hardly be used for self-stabilization where liveness must be achieved from any state in an infinite-state space. Moreover, the actions of the Collatz program include division, and cannot be specified as Presburger formulas. Farzan  {\it et al.} \cite{farzan2016proving} present the technique of well-founded proof spaces for verifying the progress of threads in parameterized multi-threaded programs with finite-domain variables. They propose a finite abstraction of infinite program executions, called quantified predicate automata, which captures language inclusion problem of infinite executions.

While program verification methods inspire us, they lack sufficient machinery to tackle the Collatz conjecture. Specifically, verifying the self-stabilization of Collatz program requires  convergence from every single concrete state in an infinite state space, whereas existing verification methods prove correctness from a set of initial states. Moreover,  the  operations in  the actions of Collatz program cannot be captured as Presburger formulas. Furthermore, abstraction techniques will be of little help because the domain of $x$ cannot be abstracted and convergence must be guaranteed from every concrete state/value. Finally, existing techniques for the verification and synthesis of self-stabilizing programs with unbounded variables  \cite{ebnenasir2022synthesizing}  would not apply either because it is unclear how we can capture the transitions of the Collatz program as semilinear sets, representing sets of periodic integer vectors.

\section{Conclusions and Future Work}
\label{sec:concl}

This paper presented a novel approach for the specification and generation of all natural values from where Collatz conjecture holds through exactly $j>0$ steps of applying the Collatz function (Problem \ref{probSpec}). We formulated the problem as a program, called the Collatz program, and reduced the conjecture into the self-stabilization of the Collatz program. We also defined the notion of a {\it convergence stair} (borrowed from the self-stabilization community with a slightly different definition) as the set of values from where the Collatz program converges to the set $\mathcal{I}_u$ of powers of two in exactly $j>0$ steps. The proposed specification is an over-approximation of the Collatz numbers in the $j$-th stair, which is then fine tuned through a verification step embedded in the generation algorithm. A significant impact of specifying the $j$-th stair for all $j>0$ (i.e., solving Problem \ref{probSpec}) includes the total order imposed on the chaotic orientation of Collatz orbits. Moreover, the proposed approach can shed light on similar conjectures \cite{briscese2024conjectures} as well as providing a systematic approach for automated analysis of the convergence of non-linear dynamic systems.

While one can generate all valid Collatz numbers belonging to any $j$-th stair (where $j>0$) using the algorithmic method of this paper and its implementation (available at \url{https://github.com/aebne/CollatzStairs}), another equally important problem remains open where the stair of a given natural number should be determined (Problem \ref{idenStair}). Solving this problem will help us prove the Collatz conjecture by showing  that the union of stairs is equal to the set of natural numbers.

\bibliographystyle{abbrv}
\bibliography{biblio}

\end{document}